# Incremental Construction of Compact Acyclic NFAs

## K. Sgarbas, N. Fakotakis, G. Kokkinakis

# Foreword

This paper uses a proof method similar to the one used in "Optimal Insertion in Deterministic DAWGs", in order to prove that the algorithm for incremental construction of acyclic Non-deterministic Finite-state Automata (NFA) that was first presented in "Two Algorithms for Incremental Construction of Directed Acyclic Word Graphs" creates NFAs that do not contain equivalent states. Unlike Deterministic Finite-state Automata (DFA), this property is not sufficient to ensure minimality, but still the resulting NFAs are considerably smaller than the minimal DFAs for the same languages. Experimental results are also provided, comparing the two structures using a lexicon of 230,000 words.

# Incremental Construction of Compact Acyclic NFAs

**Kyriakos N. Sgarbas, Nikos D. Fakotakis, George K. Kokkinakis**
Wire Communications Laboratory
Electrical and Computer Engineering Department
University of Patras, GR-26500 Rio, Greece
{sgarbas,fakotaki,gkokkin}@wcl.ee.upatras.gr

## Abstract

This paper presents and analyzes an incremental algorithm for the construction of Acyclic Non-deterministic Finite-state Automata (NFA). Automata of this type are quite useful in computational linguistics, especially for storing lexicons. The proposed algorithm produces compact NFAs, i.e. NFAs that do not contain equivalent states. Unlike Deterministic Finite-state Automata (DFA), this property is not sufficient to ensure minimality, but still the resulting NFAs are considerably smaller than the minimal DFAs for the same languages.

## 1 Introduction

Acyclic Finite-State Automata (FSA)[1] provide a very efficient data structure for lexicon representation and fast string matching, with a great variety of applications in lexicon building (Daciuk et al., 2000), morphological processing (Sgarbas et al., 2000b) and speech processing (Lacouture and De Mori, 1991). They constitute very compact representations of lexicons, since common word prefixes and suffixes are represented by the same transitions. This representation also facilitates content-addressable pattern matching.

[1] Some authors (e.g. Perrin, 1990; Aoe et al. 1992; Sgarbas et al., 1995) use the term DAWG (Directed Acyclic Word Graph) when referring to acyclic FSAs. However, others (e.g. Crochemore and Verin, 1997) use the same term to denote the suffix automaton of a string.

Examples of acyclic FSAs storing lexicons are shown in Fig.1. The FSAs consist of *states* and *transitions* between states. Each transition has a *label*. The words are stored as directed paths on the graph. They can be retrieved by traversing the graph from an initial state (*source*) to a terminal state (*sink*), collecting the labels of the transitions encountered. In this way, traversing the graphs of Fig.1 from the source (⊗) to the sink (⊙) we retrieve the words *dance*, *darts*, *dart*, *start* and *smart*.

There are two types of FSAs. The graph of Fig.1a is called *deterministic* (DFA) because no transitions exist that have the same labels and leave the same state. This property results to a very efficient search function. Graphs that do not have this property, like the one of Fig.1b, are called *non-deterministic* automata (NFA). NFAs are smaller than DFAs but they are a little slower to search.

DFAs are more popular for lexicon representation, especially the *minimal* ones, i.e. DFAs with the least number of states. Several algorithms are known for the construction of the minimal DFA, given a set of words (Hopcroft

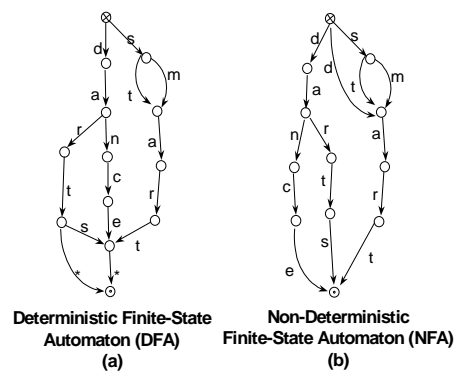

**Deterministic Finite-State Automaton (DFA)**
**(a)**

**Non-Deterministic Finite-State Automaton (NFA)**
**(b)**

Figure 1. The same lexicon in DFA and NFA.

and Ullman, 1979; Perrin, 1990; Revuz, 1992; Watson, 1993). Recently, some incremental algorithms have been proposed for this task (Aoe et al., 1993; Park et al., 1994; Sgarbas et al., 1995; Daciuk et al., 1998; Mihov, 1998; Ciura and Deorowicz, 1999; Daciuk et al., 2000; Revuz, 2000; Sgarbas et al., 2000a). Incremental algorithms are useful because they can update the lexicon without rebuilding the whole structure from scratch.[2]

For minimal NFAs there are also some (non-incremental) algorithms (Kameda and Weiner, 1970; Kim, 1974; Arnold et al., 1992; Matz and Potthoff, 1995) but unlike DFAs, there is no single minimal NFA for a given language.

In this paper we present an incremental algorithm for constructing Acyclic NFAs. We consider NFAs with one source and one sink state, like the one in Fig.1b, because this facilitates bi-directional search in the graph. We introduce the notion of a *compact* automaton[3] (i.e. one with no equivalent states) and we prove that the presented algorithm produces compact NFAs. We also show that (unlike DFAs) a compact NFA is not necessarily minimal. Therefore the algorithm does not always produce minimal acyclic NFAs; the size of the NFA depends on the order the words are inserted. However, the NFAs produced by this algorithm are typically considerably smaller than the corresponding minimal DFAs and the process of adding a new word to an existing automaton is fast enough to be used on-line.

In Section 2 of this paper some basic definitions are given that will be used throughout the paper. We have tried to define appropriate concepts that simplify the proofs of the lemmas. The presentation of the algorithm follows in Section 3 with a step-by-step example. In Section 4, a set of interesting lemmas is provided, resulting to a proof that the algorithm actually produces compact NFAs. Experimental results are presented in Section 5. The paper conclusions follow in Section 6.

---

[2]For more information, see http://odur.let.rug.nl/alfa/fsa_stuff/

[3]Not to be confused with the notion of compact DAWGs as defined by Crochemore and Verin (1997), Blumer et al. (1989), where whole strings are allowed as labels on transitions.

## 2 Definitions

Let Q be a set of *states* (vertices) and $\Sigma$ be a set of *symbols* (alphabet). A *labeled directed transition* is then defined as a triple $(n_1,n_2,s)$ from state $n_1$ to $n_2$ with label s, where $n_1,n_2 \in Q$ and $s \in \Sigma$.

Let $L \subseteq Q \times Q \times \Sigma$ be a set of labeled directed transitions. Then an ordered series $[(n_0,n_1,s_1), (n_1,n_2,s_2), (n_2,n_3,s_3), \ldots, (n_{k-2},n_{k-1},s_{k-1}), (n_{k-1},n_k,s_k)]$ of successive transitions of L is called a *succession* from state $n_0$ to state $n_k$ and is denoted by $succ(n_0,n_k)$. We say that state $n_k$ is a *successor* of state $n_0$ and that state $n_0$ is a *predecessor* of state $n_k$. In the special case where $|succ(n_0,n_k)|=1$, state $n_k$ is an *immediate successor* of $n_0$ and state $n_0$ is an *immediate predecessor* of $n_k$.

There may be more than one successions between two states. We define as $SUCC(n_0, n_k)$ the set of all successions from $n_0$ to $n_k$.

For a succession $G=succ(n_0, n_k)=[(n_0,n_1,s_1), (n_1,n_2,s_2), (n_2,n_3,s_3), \ldots, (n_{k-2},n_{k-1},s_{k-1}), (n_{k-1},n_k,s_k)]$, we define as label(G) the ordered series $[s_1, s_2, s_3, \ldots, s_{k-1}, s_k]$ of symbols as derived by the labels of the transitions in G.

For a set of successions $H=SUCC(n_0, s_k)$, we define as LABEL(H) the set of all label(G), $\forall G \in H$.

Based on the above, a *Finite State Automaton* with one source and one sink is defined as the quintuple $(Q,L,\Sigma,source,sink)$, where $source, sink \in Q$ and $\forall n \in Q-\{source\}$, $SUCC(source,n) \neq \emptyset$ and $\forall n \in Q-\{sink\}$, $SUCC(n,sink) \neq \emptyset$. In other words, *source* is a predecessor of every other state in Q and *sink* is a successor of every other state in Q.

The graph is *acyclic* iff $\forall n \in Q$, $SUCC(n,n)=\emptyset$.

The set of *strings* or *words* contained in an acyclic FSA is finite and equals to LABEL(SUCC(source,sink)). All words in an acyclic FSA are finite-lengthed.

If two states $n_1,n_2 \in Q$ (with $n_1 \neq n_2$) satisfy the property LABEL(SUCC($n_1$,sink)) = LABEL(SUCC($n_2$,sink)), then we say that $n_1$ and $n_2$ are *down-equivalent*. Accordingly, if LABEL(SUCC(source,$n_1$)) = LABEL(SUCC(source,$n_2$)), we say that $n_1$ and $n_2$ are *up-equivalent*. Saying that $n_1$ and $n_2$ are *equivalent* means that they are either up-equivalent or down-equivalent. Two states that are not equivalent are called *distinct*.

475

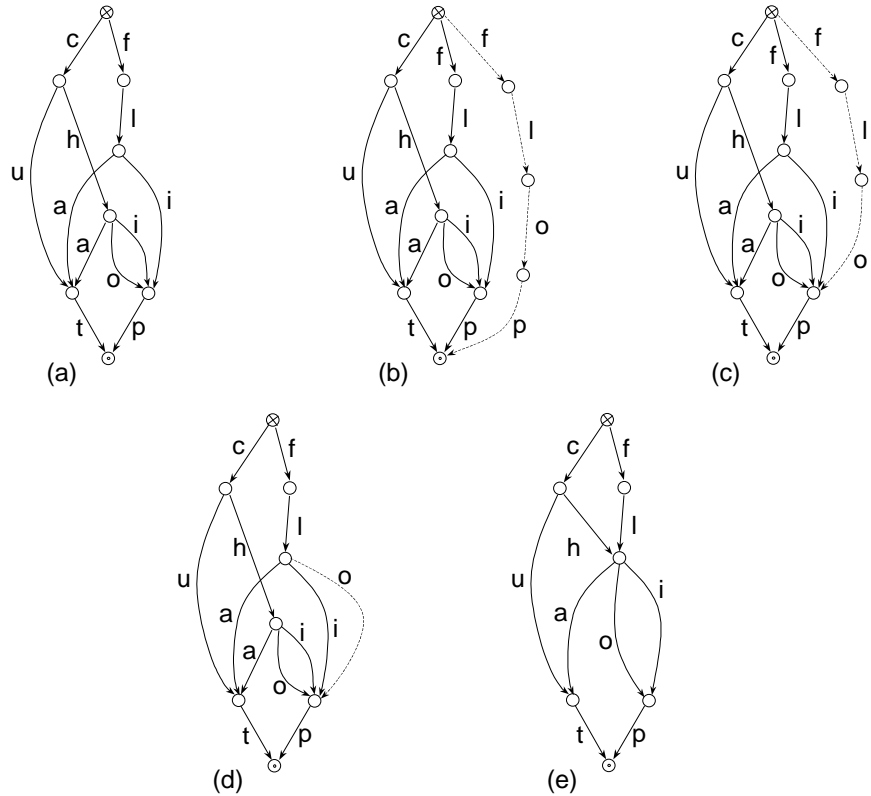

Figure 2. Operation of the incremental algorithm.

A single state is called *distinct* if it is not equivalent with any other state in Q. An automaton that contains no equivalent states is called *compact* (for DFAs compact also implies *minimal*).

For each state $n \in Q$, we consider the transitions entering n and the transitions leaving n and we define two sets: the *fan-in* set of n, $F_{IN}(n)=\{(n',s): (n',n,s) \in L\}$ and the *fan-out* set of n, $F_{OUT}(n)=\{(n',s): (n,n',s) \in L\}$. Although the transitions contained in these two sets are not full transitions (state n is missing from the triples) they are considered as transitions, since state n is always known and they can be restored into triples at any time. However, the above representation facilitates the comparison of fan-in and fan-out sets of different states.

An automaton is *deterministic* (DFA) iff $\forall n \in Q, \forall s \in \Sigma \ |\{n': (n',s) \in F_{OUT}(n)\}| \leq 1$. Thus, in DFAs $\forall n \in Q, |F_{OUT}(n)| \leq |\Sigma|$. An automaton is called *non-deterministic* (NFA) otherwise.

If two states $n_1, n_2 \in Q$ (with $n_1 \neq n_2$) satisfy the property $F_{OUT}(n_1)=F_{OUT}(n_2)$, then we say that $n_1$ and $n_2$ are *down-similar*. Respectively, if they satisfy the property $F_{IN}(n_1)=F_{IN}(n_2)$, then we say that $n_1$ and $n_2$ are *up-similar*. Saying that two states are *similar* means that they are either up-similar or down-similar. In other words, two states are similar if the input(output) transitions of the one match the input(output) transitions of the other in labels and destinations. Similar states are always equivalent, but equivalent states are not necessarily similar (see Lemma 1, below).

Let $D=(Q,L,\Sigma,source,sink)$ be an acyclic NFA as defined above. We present the following lemmas:

***Lemma 1:*** *(a) Two down-equivalent states of D are either down-similar or their immediate successors are also down-equivalent. (b) Two up-equivalent states of D are either up-similar or their immediate predecessors are also up-equivalent.*

*Proof*: (a) Let $p,q \in Q$ and p, q are down-equivalent. Consider two transitions (p,p',s) and (q,q',s) such that $p' \neq q'$. If for no $s \in \Sigma$ two such transitions exist, that implies $F_{OUT}(p)=F_{OUT}(q)$ and p,q are down-similar. Otherwise, if the transitions exist, consider the states p' and q'. Suppose that they are not down-equivalent. Thus

```
// STAGE 1: Attach word to NFA
1     n_0←source; i←0;                              // array w[] contains the new word
2     while (i<M-1) {                               // M=|w[]|
3         create new state n;
4         create new transition (n_0,n,w[i]);
5         if i=0 then k←(n,w[0]);
6         n_0←n; i++;
      }
7     create new transition (n_0,sink,w[i]);
8     j←(n_0,w[i]); p←sink; q←source;

// STAGE 2: Search for up-similar states
9     searchup_failed←0; (n',c) ←j;                 // the last-marked transition
10    while exists another (n,c)∈F_IN(p) with n≠n' {   // search only transitions in F_IN(p)
11        if F_OUT(n)=F_OUT(n') then {               // similarity check
12            {j}← F_IN(n');                         // |F_IN(n')|=1
13            F_IN(n) ←F_IN(n)∪{j};                  // redirect j
14            delete n', F_IN(n'), F_OUT(n');
15            p←n; go to 9;                          // j is the next last-marked transition
          }
      }
16    searchup_failed←1;

// STAGE 3: Search for down-similar states
17    searchdown_failed←0; (n',c) ←k;               // the first-marked transition
18    while exists another (n,c)∈F_OUT(q) with n≠n' {  // search only transitions in F_OUT(q)
19        if F_IN(n)=F_IN(n') then {                 // similarity check
20            {k}← F_OUT(n');                        // |F_OUT(n')|=1
21            F_OUT(n) ←F_OUT(n)∪{k};                // redirect k
22            delete n', F_OUT(n'), F_IN(n');
23            q←n; go to 17;                         // k is the next last-marked transition
          }
      }
24    searchdown_failed←1;

// STAGE 4: Repeat until both Stage 2 and Stage 3 fail
25    if searchup_failed+searchdown_failed < 2 then go to 9
26    end
```

Table 1. The incremental construction algorithm.

we can find two successions $s_p=succ(p',sink)$ and $s_q=succ(q',sink)$ such that $s_p≠s_q$. Then the successions $[(p,p',s)]∪s_p$ and $[(q,q',s)]∪s_q$ will also be different. But this contradicts to the assumption that p and q are down-equivalent. Therefore p' and q' are down-equivalent. (b) Symmetrical to (a). □

**Lemma 2:** *D is not compact iff similar states exist in Q.*

*Proof*: First we show that if $p,q∈Q$ and p,q are similar, then D is not compact: Similar states are always equivalent. Therefore p and q are equivalent and D is not compact. Next we show that if D is not compact then Q contains similar states: If D is not minimal then we can find $p,q∈Q$ with p and q equivalent. By Lemma 1, it is either p similar to q, or their immediate successors (predecessors) p' and q' are equivalent. Supposing the latter, we can apply Lemma 1 to p' and q'. Since there is only one sink and one source and |succ(p,sink)|, |succ(source,p)| are finite, we eventually arrive in two equivalent states which are also similar. □

Lemma 2 constitutes a very useful criterion for checking whether an automaton is compact or not. Checking using this criterion is more efficient than searching for equivalent states, since given two states, it is much faster to decide if they are similar than it is to check their equivalence.

## 3 Presentation of the Algorithm

The proposed algorithm adds a new word to an existing acyclic NFA. Figure 2 displays an example of word insertion. The original NFA of Fig.2a contains six words: *cut*, *chat*, *chop*, *chip*, *flat* and *flip*. We wish to add the word *flop*. The insertion is performed as following:

*STAGE 1*: First we attach the new word to the existing NFA by creating a separate succession of transitions between the source and the sink. We mark all these transitions. In Fig.2b they appear dashed.

*STAGE 2*: We consider the marked transition (n',p,c) closer to the sink and we search in $F_{IN}(p)$ for a state n down-similar to n'. State n' is deleted (and so all its outgoing transitions) after redirecting its incoming transition to n: $F_{IN}(n)=F_{IN}(n)\cup F_{IN}(n')$. The process is repeated, again considering the marked transition closest to the sink, until no down-similar states can be found (see Fig.2c).

*STAGE 3*: We consider the marked transition (c,p,n') closer to the source and we search in $F_{OUT}(p)$ for a state n up-similar to n'. State n' is deleted (and so all its incoming transitions) after redirecting its outgoing transition to n: $F_{OUT}(n)=F_{OUT}(n)\cup F_{OUT}(n')$. The process is repeated, again considering the marked transition closest to the source, until no up-similar states can be found (see Fig.2d).

*STAGE 4*: Stages 2 and 3 are repeated until both of them fail to find similar states (see Fig.2e).

The updated NFA contains all the words of the original one, plus the new word. Note that the algorithm does not traverse every state and transition of the original NFA to add the new word. However, the resulting NFA is compact. A proof of this is given in the next section. The algorithm is shown in Table 1.

## 4 Proof of Correctness

For the analysis of this section, consider again the example of Fig.2. The original NFA of Fig.2a is compact.

The algorithm creates new states and transitions in the first stage and it deletes states and transitions in the following stages. The NFA at the end of Stage 1 (Fig.2b) contains all the words of the original NFA plus the new word, but it is not compact.

The compaction is performed in Stages 2-4, based on the criterion of Lemma 2: Since a non-compact NFA always contains similar states, Stages 2-4 find similar states and merge them, until no more similar states can be found in the NFA. Then by Lemma 2 the NFA will be compact. Note that the algorithm does not search the whole NFA to find similar states. For every transition (n',p,c) in the path of the newly inserted word, starting from the one closest to the sink (source) and continuing upwards (downwards), it considers only the states n such that $(n,c)\in F_{IN(OUT)}(p)$ and n≠n', and it checks only them for similarity with n'.

Now consider the dashed (marked) transitions of Fig.2b. The algorithm keeps track of them. They form a succession that corresponds to the new word. Let Z be the set of states contained in that succession, excluding source and sink. Then Z contains all the new states created by the process. If any state n' in Z is found similar to some other state, state n' is deleted from Z.

Following are some interesting lemmas concerning properties of acyclic NFAs. The last one proves the correctness of the algorithm.

*Lemma 3*: *For every n∈Z, $|F_{IN}(n)|=|F_{OUT}(n)|=1$.*

*Proof*: This property is self-evident at Stage 1 where all states in Z form a succession from source to sink. During Stages 2 and 3 states and transitions are deleted from the edges of the succession only. Therefore the property is maintained throughout the whole process. □

*Lemma 4*: *(a) There is only one state n∈Z such that exists (p,n,c) with p∈Q-Z. (b) There is only one state n∈Z such that exists (n,q,c) with q∈Q-Z.*

*Proof*: (a) Suppose there are n,m∈Z and p∈Q-Z such that both (p,n,c) and (p,m,d) exist. But since n and m belong to the same succession, one must be successor of the other. Let m the successor and n the predecessor. Then there should also exist a transition (k,m,e) where k∈Z. That implies $|F_{IN}(n)|>1$, contradicting to Lemma 3. (b) Symmetrical to (a). □

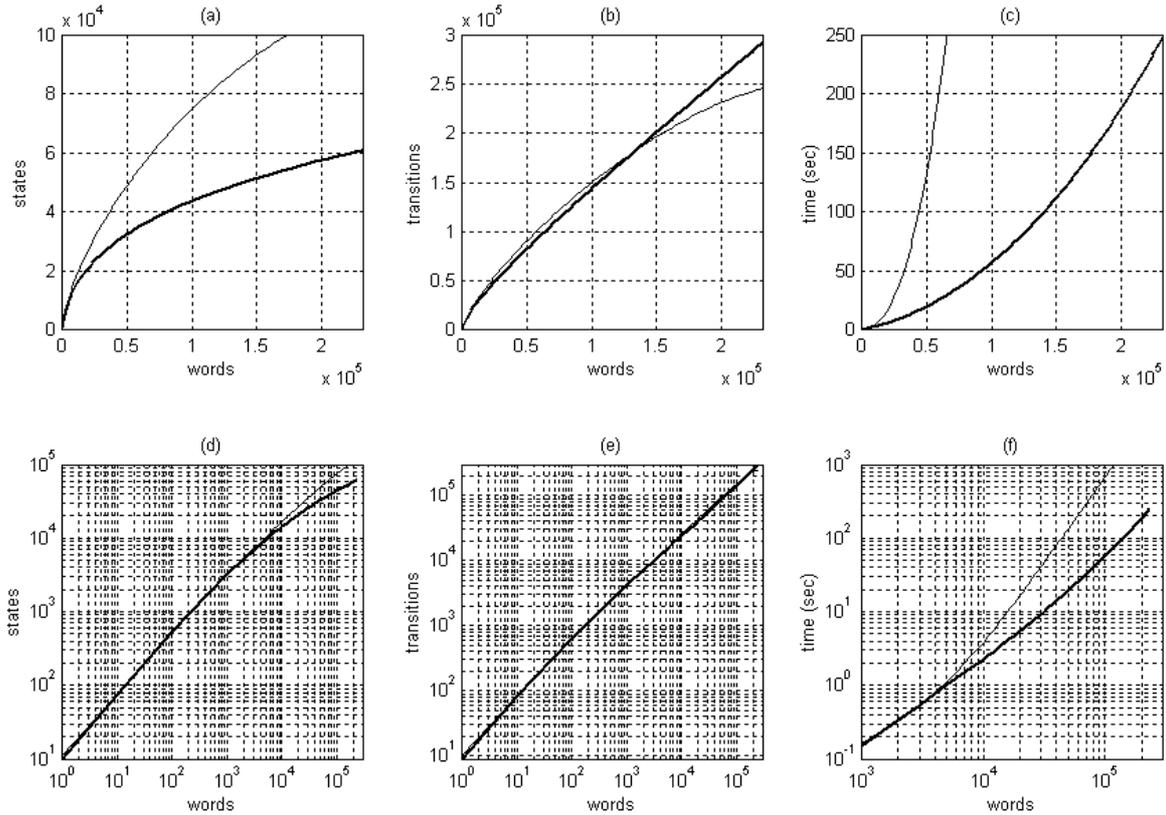

Figure 3. Test results and comparison with minimal DFA.

**Lemma 5:** *There are no equivalent states in Z.*

*Proof*: Suppose there are two equivalent states in Z. Then by Lemma 1 there are two similar states in Z. Let $p,q \in Z$ be these states. Then there are either two transitions (n,p,c) (n,q,c) or two transitions (p,n,c) (q,n,c). In either case n cannot belong to Z due to Lemma 3. Therefore $n \in Q-Z$ and by Lemma 4, p and q cannot both belong to Z. □

**Lemma 6:** *There are no equivalent states in the set Q-Z.*

*Proof*: Suppose there are two equivalent states in Q-Z. Then, by Lemma 1 we can find two similar states $p,q \in Q-Z$. Since they are similar they should be directly linked to at least one state n. By Lemma 3 n cannot belong to Z. It should belong to Q-Z. But in that case the original NFA also contains p, q and n and thus it could not be compact. □

**Lemma 7:** *Every state in Z has at most one equivalent state.*

*Proof*: Let $n \in Z$, $p,q \in Q$ such that n is equivalent to both p and q. By Lemma 5, neither p nor q can belong to Z, since $n \in Z$. Thus they must both belong to Q-Z. But since p and q are also equivalent, this contradicts Lemma 6. □

**Lemma 8:** *If we use the described algorithm to add a new word to a compact acyclic NFA, then the updated NFA is also compact.*

*Proof*: Suppose that after the end of Stage 4 the updated NFA is not compact. Then by Lemma 2 there should be two similar states in Q. Let us examine in what sets these two states can belong to: By Lemma 5 they cannot both belong to Z. By Lemma 6 then cannot both belong to Q-Z. Therefore one must belong to Z and the other to Q-Z. But since the process has been completed, Stages 2 and 3 have already checked Z for similar states. Therefore it is not possible to find two similar (or equivalent) states in Q. Thus the updated NFA is compact. □

## 5 Experimental Results

The described algorithm has been tested using a lexicon of 230,000 Greek words in random order. The average word length in the lexicon was 9.5 characters; the size of the alphabet was 36. The number of states, transitions and the construction time were measured. The results are shown in Fig.3. The thick lines refer to the NFA; the thin lines refer to the corresponding minimal DFA. For the construction of the minimal DFA an incremental algorithm was used (Sgarbas et al., 2000a) with $O(n^2)$ time performance. The test was performed on a 200 MHz PC.

Figures 3a, 3b and 3c display respectively the number of states, the number of transitions and the construction time of the automaton, in respect to the size of the lexicon (number of words). It is evident that the compact NFA constructed by the proposed algorithm had much fewer states than the corresponding minimal DFA and its construction time was notably short. However, for lexicon size grater than 130,000 words, the algorithm was less efficient concerning the number of transitions (see Fig.3b).

The same results are also shown in logarithmic scales in Figs.3d, 3e and 3f, respectively. The slopes of the lines indicate linear growth of transitions, less than linear growth of states and time complexity between $O(n)$ and $O(n^2)$.

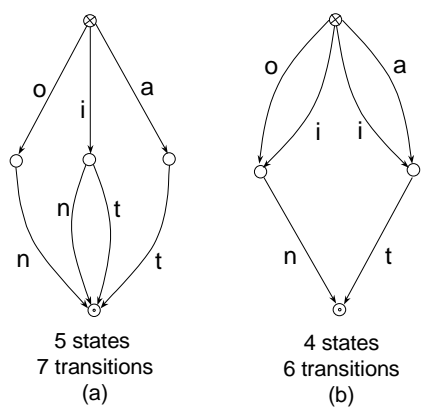

Figure 4. Insertion order affects NFA size.

## 6  Conclusion

We have presented an algorithm for adding words (strings) in acyclic NFAs and proved its compact behavior, i.e. if this algorithm is applied to a compact acyclic NFA, then the resulting NFA will also be compact. This algorithm provides an efficient and elegant way to update acyclic NFAs without having to build them from scratch. In experiments with Greek lexicons, compact NFAs constructed by the described algorithm typically required significantly less states than the corresponding minimal DFAs and about the same number of transitions, while their construction time was short enough to be used for on-line updates of lexicons. However, the proposed algorithm produces compact, but not necessarily minimal NFAs, because the order of the inserted words affects the size of the automaton. To illustrate this, consider Fig.4. Both NFAs of Fig.4 represent the same lexicon and they have both been produced by the described algorithm. However, in the case of Fig.4a the words were inserted in the order: [*in, it, at, on*], while in the case of Fig.4b the words were inserted in the order: [*in, on, at, it*]. Evidently, both NFAs are compact, but only the one of Fig.4b is minimal.

Apart from its theoretic interest, this algorithm has direct practical uses. On-line word insertion is highly desirable in every application where the data need to be updated regularly (e.g. spell-checkers) and the size of the structure is important.

## Acknowledgements

The work presented in this paper was part of the R&D project DELOS (ΕΠΕΤ-ΙΙ, 98ΓΤ-12), funded by the Greek Ministry of Development, General Secretariat of Research and Technology.

The authors would like to thank the members of FSA-Research@yahoogroups.com, especially Gertjan van Noord, Mark-Jan Nederhof, Jan Daciuk, Dale Gerdemann and Bruce Watson for their discussions on NFA minimization and Burak Emir for his help on porting the AMoRE program to Linux.

## References

J. Aoe, K. Morimoto and M. Hase. 1993. An Algorithm for Compressing Common Suffixes Used in Trie Structures. *Systems and Computers in Japan,* 24(12):31-42 (Translated from *Trans. IEICE,* J75-D-II(4):770-779, 1992).

A. Arnold, A. Dicky, M. Nivat. 1992. A Note About Minimal Non-deterministic Automata. *Bulletin of the EATCS,* 47:166-169.

A. Blumer, D. Haussler, and A. Ehrenfeucht. 1989. Average Sizes of Suffix Trees and DAWGs. *Discrete Applied Mathematics,* 24:37-45.

M. Ciura and S. Deorowicz. 1999. Experimental Study of Finite Automata Storing Static Lexicons. Report BW-453/RAu-2/99 (Also at http://www-zo.iinf.polsl.gliwice.pl/~sdeor/pub.htm).

M. Crochemore and R. Verin. 1997. Direct Construction of Compact Directed Acyclic Word Graphs. $8^{th}$ *Annual Symposium, CPM 97,* Aarhus, Denmark, 116-129.

J. Daciuk, S. Mihov, B. Watson and R. Watson. 2000. Incremental Construction of Minimal Acyclic Finite State Automata. *Computational Linguistics,* 26(1):3-16.

J. Daciuk, R.E. Watson and B.W. Watson. 1998. Incremental Construction of Acyclic Finite-State Automata and Transducers. *Proceedings of Finite State Methods in Natural Language Processing,* Bilkent University, Ankara, Turkey.

J. E. Hopcroft and J. D. Ullman. 1979. *Introduction to Automata Theory, Languages, and Computation.* Addison-Wesley, USA.

T. Kameda, P. Weiner. 1970. On the State Minimization of Nondeterministic Finite Automata. *IEEE Trans. Comp.,* C-19:617-627.

J. Kim. 1974. State Minimization on Nondeterministic Machines. IBM T. J. Watson Res. Center, Rep. RC 4896.

R. Lacouture and R. De Mori. 1991. Lexical Tree Compression. *EuroSpeech '91, $2^{nd}$ European Conference on Speech Communications and Techniques,* Genova, Italy, 581-584.

O. Matz, A. Miller, A. Potthoff, W. Thomas and E. Valkema. 1995. Report on the Program AMoRE. Bericht 9507, Institut fur Informatik und Praktische Mathematik, Christian-Albrechts-Universitat zu Kiel (Also at ftp://ftp.informatik.uni-kiel.de/pub/kiel/amore).

O. Matz and A. Potthoff. 1995. Computing Small Nondeterministic Finite Automata. *Proc. Workshop on Tools and Algorithms for the Construction and Analysis of Systems,* Dept. of CS, Univ. of Aarhus, 74-88.

S. Mihov. 1998. Direct Construction of Minimal Acyclic Finite States Automata. *Annuaire de l' Universite de Sofia ' St. Kl. Ohridski'',Faculte de Mathematique et Informatique,* Sofia, Bulgaria, 92(2).

K. Park, J. Aoe, K. Morimoto and M. Shishibori. 1994. An Algorithm for Dynamic Processing of DAWG's. *International Journal of Computer Mathematics,* Gordon and Breach Publishers SA, OPA Amsterdam BV, 54:155-173.

D. Perrin. 1990. Finite Automata. In: J. van Leeuwen, ed., *Handbook of Theoretical Computer Science,* Elsevier, Amsterdam, Vol. A, 3-57.

D. Revuz. 1992. Minimization of Acyclic Deterministic Automata in Linear Time. *Theoretical Computer Science,* Elsevier 92:181-189.

D. Revuz. 2000. Dynamic Acyclic Minimal Automaton. CIAA 2000, $5^{th}$ International Conference on Implementation and Application of Automata, London, Canada pp.226-232.

K. Sgarbas, N. Fakotakis and G. Kokkinakis. 1995. Two Algorithms for Incremental Construction of Directed Acyclic Word Graphs. *International Journal on Artificial Intelligence Tools*, World Scientific, 4(3):369-381.

K. Sgarbas, N. Fakotakis and G. Kokkinakis. 2000a. Optimal Insertion in Deterministic DAWGs. Technical Report WCL/SLT#000524, Wire Communications Lab., Dept. of Electrical and Computer Engineering, University of Patras, Greece (Also at http://slt.wcl.ee.upatras.gr/sgarbas/PublAbsEN.htm).

K. Sgarbas, N. Fakotakis and G. Kokkinakis. 2000b. A Straightforward Approach to Morphological Analysis and Synthesis. *Proc. COMLEX 2000, Workshop on Computational Lexicography and Multimedia Dictionaries,* Kato Achaia, Greece, 31-34.

B. Watson. 1993. A Taxonomy of Finite Automata Minimization Algorithms. Computer Science Note 93/44, Eindhoven University of Technology, The Netherlands (Also at http://www.OpenFIRE.org).